\documentclass[longauth,letter]{aa} % for the letters 
\usepackage{graphicx}
\usepackage{txfonts}
\begin{document} 

   \title{Visible and near-infrared observations of asteroid 2012 DA$_{14}$\\
	during its closest approach of February 15, 2013}

   \author{J. de Le\'on\inst{1,5}
          \and
          J. L. Ortiz\inst{2}
				  \and
	        N. Pinilla-Alonso\inst{3}
	        \and
	        A. Cabrera-Lavers\inst{4,5,6}
	        \and
	        A. Alvarez-Candal\inst{2,7}
					\and
		      N. Morales\inst{2}
					\and
					R. Duffard\inst{2}
					\and
					P. Santos-Sanz\inst{2}
					\and
					J. Licandro\inst{5,6}
					\and
					A. P\'erez-Romero\inst{4}
					\and 
					V. Lorenzi\inst{8}
					\and
					S. Cikota\inst{9}
          }

   \institute{Departamento de Edafolog\'ia y Geolog\'ia, Universidad de La Laguna (ULL), Avda. Astrof\'isico Francisco S\'anchez, s/n, E-38205, La Laguna, Tenerife, Spain, \email{julia.de.leon@ull.es}\\
         \and
				      Instituto de Astrof\'isica de Andaluc\'ia - CSIC, Glorieta de la Astronom\'ia s/n, E-18008, Granada, Spain\\
				 \and
              Earth and Planetary Sciences Department, University of Tennessee, Knoxville, TN 37996, USA\\
			   \and
					    GTC Project, E-38205, La Laguna, Tenerife, Spain\\
				 \and
				      Instituto de Astrof\'isica de Canarias (IAC), C/V\'ia L\'actea s/n, E-38205, La Laguna, Spain\\
				 \and
				      Departamento de Astrof\'isica (ULL), E-38205, La Laguna, Spain\\
				 \and
				      European Southern Observatory, Alonso de C\'ordova 3107, Vitacura, Casilla 19001, Santiago 19, Chile\\
					\and		
							Fundaci\'on Galileo Galilei - INAF, Rambla Jos\'e Ana Fern\'andez P\'erez 7, 37812, La Palma, Spain\\
					\and
					    Physics Department, University of Split, Nikole Tesle 12, 21000 Split, Croatia\\
             }

   \date{Received February 27, 2013; accepted June 17, 2013}

% \abstract{}{}{}{}{} 
% 5 {} token are mandatory
 
  \abstract
   {Near-Earth asteroid 2012 DA$_{14}$ made its closest approach on February 15, 2013, when it passed at a distance of 27,700 km from the Earth's surface. It was the first time an asteroid of moderate size was predicted to approach that close to the Earth, becoming bright enough to permit a detailed study from ground-based telescopes.}
   {Asteroid 2012 DA$_{14}$ was poorly characterized before its closest approach. The main objective of this work was to obtain new and valuable data to better understand its physical properties, and to evaluate the effects of such a close approach on the object.}
   {We acquired data using several telescopes on four Spanish observatories: the 10.4m Gran Telescopio Canarias (GTC) and the 3.6m Telescopio Nazionale Galileo (TNG), both in the El Roque de los Muchachos Observatory (ORM, La Palma); the 2.2m CAHA telescope, in the Calar Alto Observatory (Almer\'ia); the f/3 0.77m telescope in the La Hita Observatory (Toledo); and the f/8 1.5m telescope in the Sierra Nevada Observatory (OSN, Granada). We obtained visible and near-infrared color photometry, visible spectra and time-series photometry.}
   {Visible spectra together with visible and near-infrared color photometry of 2012 DA$_{14}$ show that the object can be classified as an L-type asteroid, a rare spectral type among the asteroid population, with a composition similar to that of carbonaceous chondrites. The time-series photometry provides a rotational period of 8.95 $\pm$ 0.08  hours after the closest approach, and there are indications that the object suffered a spin-up during this event. The large amplitude of the light curve suggests that the object is very elongated and irregular, with an equivalent diameter of around 18m. We obtain an absolute magnitude of H$_R$ = 24.5 $\pm$ 0.2, corresponding to H$_V$ = 25.0 $\pm$ 0.2 in V. The GTC photometry also gives H$_V$ = 25.29 $\pm$ 0.14. Both values agree with the value listed at the Minor Planet Center (MPC) shortly after discovery, although H$_V$ is very sensitive to the slope parameter G used to correct for phase angle. From the absolute photometry, together with some constraints on size and shape, we compute a geometric albedo of $p_V$ = 0.44 $\pm$ 0.20, which is slightly above the range of albedos known for L-type asteroids (0.082 - 0.405).}
   {}

   \authorrunning{de Le\'on et al.}
   \titlerunning{Observations of asteroid 2012 DA$_{14}$}
   \maketitle
%
%________________________________________________________________

\section{Introduction}\label{intro}
 Asteroid 2012 DA$_{14}$ (hereafter DA14) is a near-Earth object discovered on February 23 at La Sagra Observatory, Spain\footnote{http://www.minorplanetcenter.net/mpec/K12/K12D51.html}. This Apollo near-Earth asteroid focused the attention of the scientific community when it made its closest approach to the Earth on February 15, 2013, at 19:26 UT. It passed at a distance of 27,700 km from the Earth's surface, inside the geosynchronous satellite ring. Asteroid DA14 is poorly characterized. Prior to the closest approach there was information scattered on the internet (possible L-type and a rotation period $\sim$6 hours), but this information had not been published in any scientific publication. According to the value listed by the MPC, it has an absolute magnitude of H = 24.4 and a diameter of $\sim$50m, assuming a geometric albedo of 20\%, which is an average albedo for asteroids. This introduces an error in diameter of around a factor of two, and an even larger uncertainty in its mass estimation. 

\begin{table*}
\caption{Observational details of the color photometry and spectroscopy data. All data were obtained with a distance to the Sun $r = 0.988$ AU. The UT corresponds to the middle exposure time and in all cases refers to February 16. }\label{table1}      
\centering
\small          
\begin{tabular}{l c c c c c c c | c c c c c }     % 13 columns 
\hline\hline\\[-2.5mm]       
\multicolumn{8}{l}{{\bf Photometry}} & \multicolumn{5}{l}{{\bf Spectroscopy}}\\
\hline\\[-2.5mm]           
Filter  &  $\lambda_c$ ($\mu$m)  &  UT  & Exp.  (secs) & Airm. & $\Delta$ (AU)  & $\alpha$ ($^{\circ}$) & Magnitude  & UT & Exp. (secs) & Airm.& $\Delta$ (AU)  & $\alpha$ ($^{\circ}$) \\
\hline\\[-2.5mm]
\multicolumn{8}{l}{OSIRIS-GTC} & \multicolumn{5}{l}{CAFOS-CAHA}\\			
\hline\\[-2.5mm]									
   $g'$ & 0.48 & 01:08 & 1     & 1.86 &  0.00093 & 73.09 & 12.22 $\pm$ 0.02 & 00:40 & 60 & 1.45 & 0.00085 & 72.00\\
           &           & 05:18  & 4x5 & 1.90 &  0.00156 & 77.68 & 14.21 $\pm$ 0.03 & 00:44 & 60 & 1.45 & 0.00087 & 72.17 \\
 $ r'$ & 0.64 & 01:07 & 1      & 1.86  & 0.00092 & 73.06 & 11.50 $\pm$ 0.02 &  00:47 & 60&  1.45 & 0.00088 & 72.29 \\
	        &           & 05:20 & 4x1 & 1.90   & 0.00157 & 77.70 & 13.59 $\pm$ 0.03 & 00:50 & 60 & 1.45  & 0.00088 & 72.42 \\
  $i'$ & 0.77 & 01:08 & 1      & 1.86  & 0.00093 & 73.09 & 11.32 $\pm$ 0.03 &  02:37 & 120 & 1.49  & 0.00115 & 75.51 \\
          &           & 05:22 &  4x1 & 1.90 &  0.00157 & 77.71 & 13.44 $\pm$ 0.03 & 02:41 &  120& 1.49& 0.00116 & 75.59 \\
   $z'$ & 0.96 & 01:09 & 1 & 1.86 & 0.00093 & 73.13 & 11.22 $\pm$ 0.03   &            &        &           &                  &            \\
         &           & 05:25 & 4x3	& 1.90 & 0.00158 & 77.74 & 13.37 $\pm$ 0.02 &            &        &           &                  &            \\	
\hline\\[-2.5mm]
\multicolumn{8}{l}{NICS-TNG} & \multicolumn{5}{l}{OSIRIS-GTC}\\
\hline\\[-2.5mm]
1mic & 1.02 & 05:32 & 10x10 & 1.90 & 0.00160 & 77.79 & 12.69 $\pm$ 0.04 & 05:30  &  90 & 1.90   & 0.00159 & 77.78 \\
$J$ & 1.27 & 05:37 & 10x10 & 1.90 & 0.00161	 & 77.83 & 12.47 $\pm$ 0.06 &            &        &           &                  &            \\
$H$ & 1.60 & 05:42 & 10x10 & 1.90 & 0.00162 & 77.87 & 12.25 $\pm$ 0.06 &            &        &           &                  &               \\ 
$K$ & 2.2 & 05:47 & 10x10 & 1.90 & 0.00163 & 77.91 & 12.29 $\pm$ 0.08 &            &        &           &                  &             \\ 
\hline                  
\end{tabular}
\end{table*}

Close approaches of asteroids to the Earth are interesting events because the objects can potentially break up due to tidal forces if they reach the Roche limit. They can even be a mechanism for the formation of binaries (e.g. Walsh \& Richardson \cite{walsh06}) and perhaps asteroid pairs. In general, interchanges of linear and angular momentum between the Earth and the approaching body are processes that have an effect on the population of near-Earth objects by altering their orbits and their physical properties. For these reasons, we took advantage of DA14's closest approach to Earth on February 15, 2013 to obtain highly complementary data that night (visible and near-infrared color photometry, visible spectroscopy, and time-series photometry) using the telescopes listed in the abstract.
%__________________________________________________________________
\section{Observations and data reduction}\label{obs}

Visible broad-band photometry was obtained using the Optical System for Imaging and Low Resolution Integrated Spectroscopy (OSIRIS) camera-spectrograph (Cepa et al. \cite{cepa00}) at the GTC. The OSIRIS instrument consists of a mosaic of two Marconi CCD detectors, each with 2048x4096 pixels and a total field of view of 7.8'x7.8' (plate scale of 0.127 ''/pix). The CCD was binned in 2x2 pixels. Observations were made under photometric conditions, dark moon, and an average seeing of 2.0". Two different series in Sloan $g', r', i'$, and $z'$ filters were obtained. Observational details are summarized in Table \ref{table1}, where we show the mid-exposure UT time of each observation, the distance to the Earth ($\Delta$), and the phase angle ($\alpha$). Bias correction, flat-fielding and bad-pixel-masking were done using standard procedures, and the images were finally aligned to perform the photometry. Images were calibrated using Sloan photometric standards and average extinction coefficients. The obtained magnitudes on each filter are shown in Table \ref{table1}. 
	  
Broad-band photometry in the near-infrared was performed using the NICS camera-spectrograph (Baffa et al. \cite{baffa01}) at the TNG. The plate scale was 0.25 ''/pixel, yielding a field of view of 4.2'x4.2'. The series of images through a custom filter centered at 1.02 $\mu$m (1mic) and the standard Johnson $J$, $H$, and $K$ filters consisted of ten individual exposures of 10 secs following a dithering pattern on different positions on the CCD, separated by offsets of 10 pixels. The tracking of the telescope was at the proper motion of the target. Each frame was first corrected for cross-talk, then the data were  reduced in the standard way using IRAF routines. All frames were flat-field corrected and sky subtracted. Standard aperture photometry was done. We observed two fields of standard stars for calibration, using a mosaic of five individual exposures of 10 secs for P177-D and 5 secs for P041-C (Persson et al. \cite{persson98}). The resulting magnitudes are shown in Table \ref{table1} together with the observational details. 
		
A visible spectrum of DA14 in the 0.49--0.92 $\mu$m range was obtained with OSIRIS-GTC using the R300R grism and a 10'' slit to minimize possible slit losses due to atmospheric dispersion. The R300R grism provides a dispersion of 7.64 \AA/pixel. Observational details are shown in Table \ref{table1}. Images were bias- and flat-field corrected, using lamp flats. The two-dimensional spectra were wavelength calibrated using Xe+Ne+HgAr lamps. After the wavelength calibration, sky background was subtracted and a one-dimensional spectrum was obtained. To correct for telluric absorption and to obtain the relative reflectance, the solar-analog star SA110-361 (Landolt \cite{landolt92}) was observed using the same spectral configuration at an airmass nearly identical to that of the object. The spectrum of the object was then divided by the corresponding spectrum of the solar analog, and normalized to the central wavelength of the $r'$ filter (0.64 $\mu$m). The resulting final reflectance spectrum is shown in red in Fig. \ref{figure1}.

Another series of visible spectra of DA14 in the range 0.45--0.92 $\mu$m were obtained on the same night using the Calar Alto focal reducer and faint object spectrograph CAFOS instrument. CAFOS is equipped with a 2048x2048 pixel CCD and a plate scale of 0.52''/pixel. We used the R-400 grism with a dispersion of 9.7\AA/per pixel ($R \sim 400$). A 2'' slit was employed. The tracking was at the asteroid's proper motion and the slit was oriented in the direction of asteroid's motion to minimize slit losses. Six individual spectra were obtained (see details in Table \ref{table1}). Pre-processing of the CCD images included bias and flat-field correction. A one-dimensional spectrum was extracted from two-dimensional images, and wavelength calibration was applied using Cd, Hg, and Rb lamps. All spectra were corrected for atmospheric extinction. To obtain the asteroid's reflectance spectra we observed the solar-analog star BS4486. The final visible spectrum is the average of the six individual spectra and normalized to unity at the central wavelength of the $r'$ filter (black line in Fig. \ref{figure1}).

Time-series photometry was obtained from CCD images acquired at the remotely operated f/3 0.77m telescope in the La Hita Observatory. The images were obtained using exposure times of 1 and 2 secs with a 4k x 4k CCD that provides a field of view of 47.9'x47.9' and using a Luminance filter. Binning 2x2 was used, yielding a scale of 1.42 ''/pixel. The telescope was tracked at sidereal rate. The object was observed in 29 different star fields.  The observations spanned $\sim$9 hours, starting about 2 hr after the closest approach. There was a 1.3 hour gap in the acquisition due to a network problem. More than 1600 images were processed. The images were dark subtracted and flat-fielded. The flux of the asteroid was obtained through synthetic apertures, changed along the image sequence to accomodate to the different trail lengths.  Twenty-four comparison stars were measured in each image, with identical apertures to that used for the asteroid, to monitor possible atmospheric transparency fluctuations. Zeropoints for the photometry were obtained for each field using USNO-B1 stars. The magnitudes were corrected for the changing heliocentric ($r$) and geocentric ($\Delta$) distance of DA14. We also applied the phase-angle correction using Bowell's G,H formalism (Bowell et al. \cite{bowell02}). The slope parameter G simply describes the shape of the ($H$ vs. $\alpha$) data points, i. e., how the absolute magnitude $H$ changes with the phase angle. This correction is important because of the significant change during the observations (50$^{\circ}$ $< \alpha <$ 75$^{\circ}$). The resulting light curve is shown with red crosses in Fig. \ref{figure2}. A complementary set of 60 images of 10  secs were obtained using the same procedure in the following night (green crosses in Fig. \ref{figure2}).

A set of 184 images of 1 sec of exposure time obtained with the f/8 1.5m telescope at the OSN were also analyzed. The images were taken by means of a 2k x 2k CCD, in 2x2 binning mode (resolution of 0.46 ''/pixel) and using no filters. The field of view was 7.8'x7.8'. Observations started $\sim$20 minutes before dawn, and during that time the object remained in the same star field, so no re-pointing was necessary. The image processing and data reduction were identical to that used for the La Hita f/3 0.77m images and described above. The photometry was obtained with respect to USNO-B1 standard stars. Results are shown as blue crosses in Fig. \ref{figure2}.

\section{Results and discussion}\label{results}
\begin{figure}
   \centering
     \includegraphics[width=7cm]{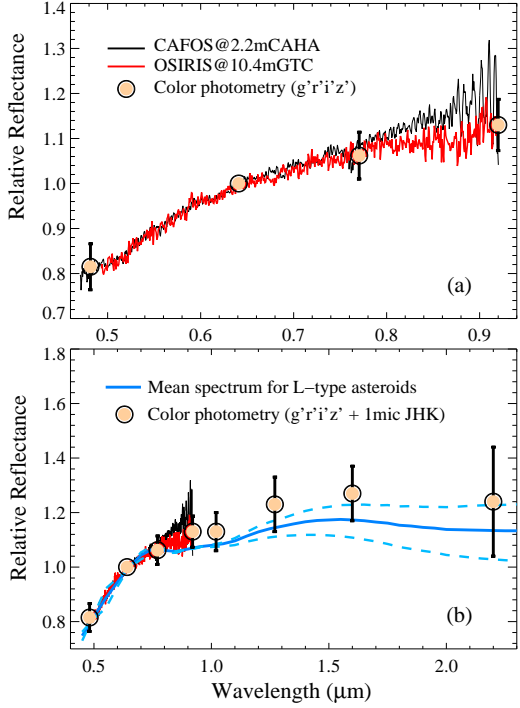}
      \caption{(a) Visible spectra of DA14 obtained with CAFOS (black) and OSIRIS (red). Filled circles are the reflectance $R$ values. (b) Same as the top panel, but adding the $R$ values for the near-infrared.  The mean spectrum of L-type asteroids from DeMeo et al. (\cite{demeo09}) is shown in blue. Dashed lines indicate the variation range of this mean spectrum. Both the spectra and the reflectance $R$ are normalized at the central wavelength of $r'$.}\label{figure1}
   \end{figure}

Using the visible and near-infrared colors, we computed the spectral reflectance $R$, normalized to the $r'$ filter. We used Sloan colors of the Sun from Sparke \& Gallagher (\cite{sparke07}) and Fukugita et al. (\cite{fukugita11}). For the filters in the near-infrared, we used Sun colors from Campins et al. (\cite{campins85}) and Colina et al. (\cite{colina96}). We also corrected $R$ for the changing distance of the asteroid and the change in phase angle, as well as the variation in magnitude due to the rotation of the object. The computed $R$ values with their corresponding error bars are shown as filled circles in Fig. \ref{figure1}, superimposed on the two visible spectra from GTC (red) and CAHA (black) in the top panel. The photometric results agree very well with the two spectra. 

\begin{figure}
   \centering
     \includegraphics[width=7cm]{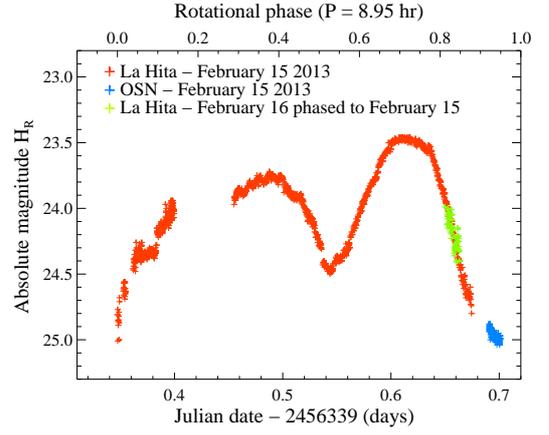}
      \caption{Resulting lightcurve from the time-series photometry made with the f/3 0.77m telescope at the La Hita Observatory  and the f/8 1.5m telescope at the Sierra Nevada Observatory.}\label{figure2}
   \end{figure}

We used the M4AST online tool (Popescu et al. \cite{popescu12}) to classify our visible spectra of DA14, obtaining a taxonomic classification as an L-type. The computed $R$ values for the near-infrared are shown in the bottom panel of Fig. \ref{figure1}, together with the mean spectrum of L-type asteroids (blue line) from DeMeo et al. (\cite{demeo09}). Light-blue dashed lines indicate the variation range of this mean spectrum. Considering the error bars, our near-infrared colors are consistent with the taxonomy obtained from the visible spectra, showing the expected behavior for L-type asteroids: a strongly reddish spectrum shortward of 0.8 $\mu$m, and a featureless flat spectrum longward of this, with little or no concave-up curvature related to a 1 $\mu$m silicon absorption band. There is often a gentle concave-down curvature in the infrared with a maximum around 1.5 $\mu$m, and there may or may not be a 2 $\mu$m absorption feature. There are almost no specific references in the literature regarding these objects, as L-type asteroids are relatively rare, accounting only for the 3-6\% of all the spectrally classfied asteroids (Bus \& Binzel \cite{bus02}; Binzel et al. \cite{binzel04}; DeMeo et al. \cite{demeo09}). The majority of the objects classified as L-types belong to two dispersed groups or families in the outer main belt, the Henan and the Watsonia families. Burbine et al. (\cite{burbine92}) analyzed the visible and near-infrared spectra of two members of the Watsonia group, finding that their spectra were most likely produced by spinel, an aluminium-magnesium oxide mineral commonly present in inclusions in CV3 and CO3 meteorites,  two groups of carbonaceous chondrites. Carbonaceous chondrites are chemically primitive (their abundances of major, non-volatile elements approach those observed in the Sun) and, with a few exceptions, they are aqueously altered or weakly thermally metamorphosed.  In this sense, L-types would not follow the well-known relation between the S-types and ordinary chondrites. 
 
Regarding DA14's light curve, the variations in Fig. \ref{figure2} are large and due to the asteroid's rotation, implying a very elongated object (albedo variations on the surface and shadowing effects can also contribute). The amplitude of the variation is large, but not unusual for small objects in the size range of DA14, as can be seen in the asteroids light curve database (Warner et al. \cite{warner09}). We would expect two maxima and two minima in the light curve during a complete rotation of an elongated object. As the two maxima we see in Fig. \ref{figure2} are very different, this asteroid must be irregular, having an axial ratio of $\sim$10$^{(\Delta \rm{m/2.5})}$= 4 (with $\Delta$m$\sim$1.5 mag).   If the long axis is around 40m, as can be inferred from the radar images shown by NASA (http://www.jpl.nasa.gov/news/), the short axis would be $\sim$10m. Owing to the shadowing effects and albedo variations, the short axis could be slightly longer. Assuming a size of 40x12x12m, the asteroid's equivalent-volume diameter would be around 18m, much shorter than the diameter estimated before the closest approach. All this suggests that approaches of objects like DA14 within the geosynchronous satellite ring are more frequent (four times a year, following Ortiz et al. \cite{ortiz06}; once a year according to Brown et al. \cite{brown02}) than estimated for an object with a diameter of 50m (once every 40 years).  For the absolute magnitude we obtained a value of H$_R$ = 24.5 $\pm$ 0.2, which translates into H$_V$ = 25.0 $\pm$ 0.2 considering the $V$-$R$ color. The error in H$_R$ comes from the comparison between USNO-B1 with Landolt standards, and the unknow uncertainty associated to the G=0.3 value we used for L- type asteroids, as there are only a few values published in the literature for this spectral class (Pravec et al. \cite{pravec12}). Aditionally, the GTC photometry allowed us to obtain an independent estimate of H$_V$=25.29 $\pm$ 0.14. Using the absolute magnitude and the estimated size of the asteroid, we compute an albedo of $p_V$ = 0.44 $\pm$ 0.20. This value is slightly above the range of albedos for L-class objects (0.082-0.405) given in Mainzer et al. (\cite{mainzer11}) for a sample of 72 L-types observed with WISE. According to these authors, very high albedo objects are suspected to have a G value different from the one typically assumed for asteroids (G=0.15). 

The rotation period of DA14 was precisely determined from the light curve obtained on the night of the closest approach by adding a small set of data that we obtained on the subsequent night (green crosses in Fig. \ref{figure2}). We searched for the best rotation period that resulted in a good match of these two datasets, obtaining 8.95 $\pm$ 0.08 hr. This value agrees with the estimate of between 8 and 9 hours from radar images after the closest approach that was reported by Lance Benner to the NASA/JPL Asteroid Radar Research. 
Our light curve at its initial part has some discontinuities and features that are probably artifacts (the object was moving more rapidly so it was difficult to obtain accurate photometry), but the overall increase of brightness is correct. We used the photometry reported to the MPC by observers around the world (MPCAT-OBS) to investigate whether we could obtain other estimates of the rotation period. Even though the photometry from MPCAT-OBS is usually very poor, we selected only the $R$-band data; the $R$-band is the most reliable of the bandpasses. The scatter in the data (after correcting for changing geo- and heliocentric distances) was smaller than the large amplitude of the rotation light curve, and so periodic features can be detected. We obtained Lomb periodograms (Lomb \cite{lomb76}) of the photometry residuals (observed values minus the expected magnitudes from the JPL Horizons service). The Lomb periodogram of unevenly sampled data is a powerful means to study the rotation of small solar system bodies (e.g. Sheppard et al. \cite{sheppard08}). We separated the residuals into two groups, one before the closest approach and another one after the closest approach. The Lomb periodogram of the residuals after the closest approach (Appendix A, Fig\ref{figure3}) has its strongest peak at 4.40 $\pm$ 0.05 hr, which corresponds to a rotation period of 8.8 $\pm$ 0.1 hr (the lightcurve is double peaked, with two maxima and two minima per rotation period). The normalized spectral power of this period is very high, well above the 99.99\% significance level. The rotation period perfectly agrees with the value derived from our own data. Therefore, this is a validation that the MPC data are good enough to reveal a periodicity for this particular object.

We obtained two peaks of similar spectral power for the group of residuals prior to the close approach, with significance levels above 99.99\%. They correspond to rotation periods of 9.8 $\pm$ 0.1 hr and 12.4 $\pm$ 0.1 hr (Appendix A, Fig\ref{figure4}). The former is a more plausible value as it requires a smaller change with respect to the post close approach value of 8.8 hr. Even though the rotation period prior to the close approach is not unambiguously determined from the data, it seems that the close approach caused a detectable spin-up on DA14. Detailed descriptions of the periodogram analysis and the implications of a change in rotation on the physical properties of DA14 are the subject of a forthcoming paper (Ortiz et al. 2013 in preparation).

\begin{acknowledgements}
 JdL acknowledges financial support from the Spanish Secretar\'ia de Estado de Investigaci\'on, Desarrollo e Innovaci\'on (Juan de la Cierva contract). JLO acknowledges support from the project AYA2011- 30106-C02-01 (MINECO). AAC acknowledges support from the Marie Curie Actions of the European Commission (FP7-COFUND). RD acknowledges financial support from the MINECO (contract Ram\'on y Cajal). JL acknowledges support from the projects AYA2011-29489-C03-02 and AYA2012-39115-C03-03 (MINECO). Based on observations made with the GTC and the TNG, located at the Spanish ORM of the Instituto de Astrof\'isica de Canarias. Based on observations collected at the Calar Alto Observatory, jointly operated by the MPIA Heidelberg and the Instituto de Astrof\'isica de Andaluc\'ia (CSIC). We thank Rene Rutten and Pedro \'Alvarez for the access to the data of 2012 DA$_{14}$ obtained with GTC. We thank E. Molinari  for allocation of director's discretionary time and the TNG's service staff. We thank Calar Alto Observatory for allocation of director's discretionary time to this programme. We thank Faustino Organero from Astrohita.
\end{acknowledgements}

\Online
\begin{appendix}
In this appendix we present the generated Lomb periodograms from the computed residuals of the photometry observations from the MPCAT-OBS. 

\begin{figure*}
\centering
\includegraphics[width=12cm, angle=90]{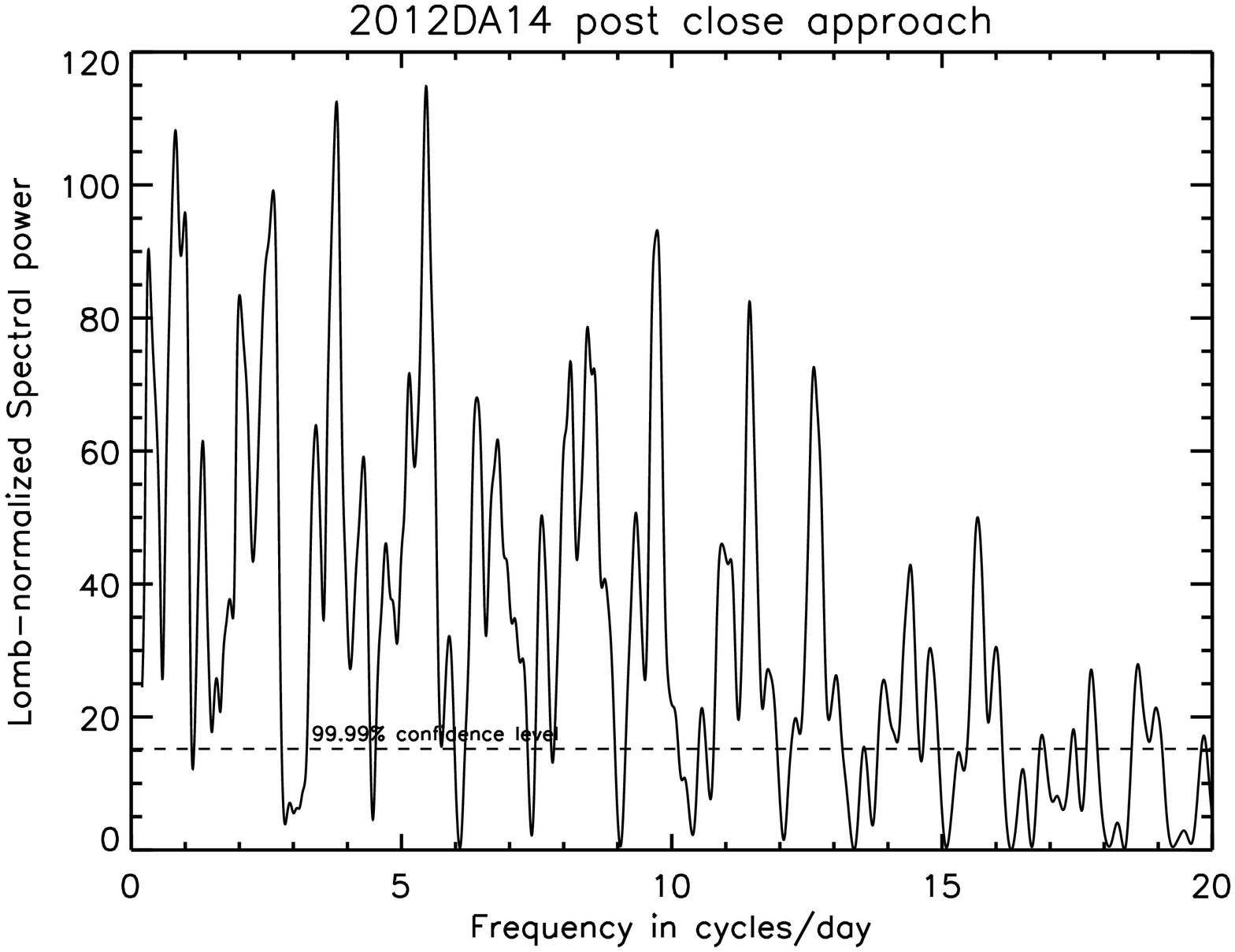}
\caption{Lomb periodogram of the MPCAT-OBS photometry residuals (observed magnitudes minus the JPL horizons expected magnitudes) after closest approach. The 99.99\% confidence level is indicated with a dashed line.}
\label{figure3}
\end{figure*}

\begin{figure*}
\centering
\includegraphics[width=12cm, angle=90]{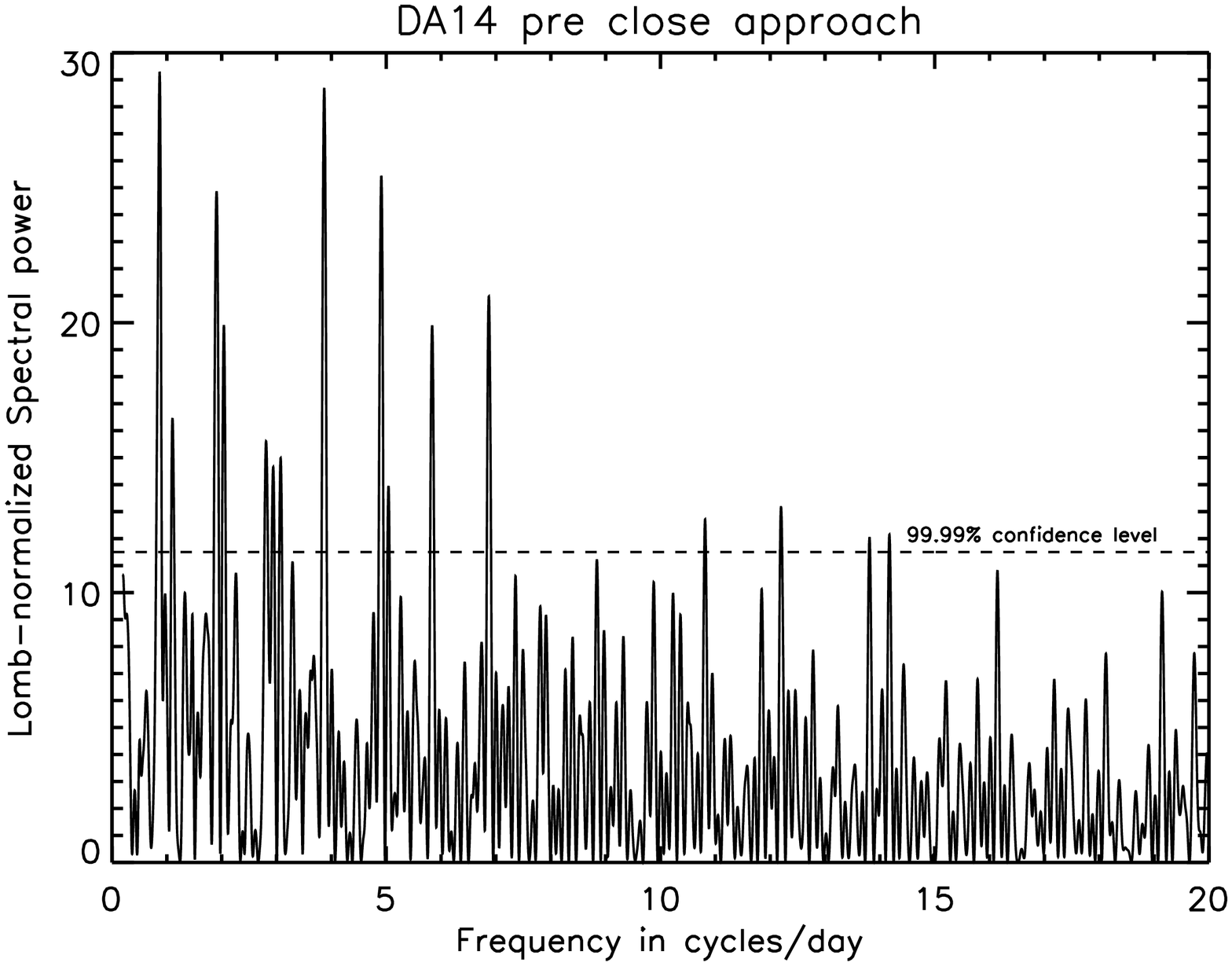}
\caption{Lomb periodogram of the MPCAT-OBS photometry residuals with respect to JPL horizons magnitudes, prior to closest approach. The 99.99\% confidence level is indicated with a dashed line.}
\label{figure4}
\end{figure*}

\end{appendix}

\end{document}